\documentclass[aps,prb,twocolumn,groupedaddress,floats]{revtex4}

\usepackage{epsfig}
\usepackage{amsmath}

\begin{document}

\renewcommand{\bottomfraction}{0.99}
\renewcommand{\topfraction}{0.99}

\draft


\title[Ferromagnetism in the Periodic Anderson Model]
      {A Dynamical Mean-Field Study of Ferromagnetism\\ in the Periodic
        Anderson Model} 




\author{D.\ Meyer}
\email[]{dietrich.meyer@physik.hu-berlin.de}
\author{W.\ Nolting}
\affiliation{Lehrstuhl Festk{\"o}rpertheorie, Institut f{\"u}r Physik,
  Humboldt-Universit{\"a}t zu Berlin, Invalidenstr.\ 110, 10115 Berlin}

\date{\today}

\begin{abstract}
The ferromagnetic phase diagram of the periodic Anderson model is
calculated using dynamical 
mean-field theory in combination with the modified perturbation
theory. Concentrating on the intermediate valence regime, the phase
boundaries are established as function of the total electron density,
the position of the atomic level and the hybridization strength. The
main contribution to the magnetic moment stems from the
$f$-electrons. The conduction band polarization is, depending on the
system parameters either parallel or antiparallel to the
$f$-magnetization. By investigating the densities of states, one
observes that the change of sign of the conduction band polarization is
closely connected to the hybridization gap, which is only apparent in
the case of almost complete polarization of the
$f$-electrons. Finite-temperature calculations are also performed, the
Curie temperature as function of electron density and $f$-level position 
are determined. In the intermediate-valence regime, the phase
transitions are found to be of second order.

\end{abstract}


\maketitle


\section{Introduction}
The periodic Anderson model (PAM) represents probably the simplest way
of modeling 
some of the rich physics found in Lanthanides and
Actinides\cite{hewson,Czy86,GS91}. One believes that
most of the physical properties typical for these materials originate from 
the interplay between the incompletely filled $4f$ or $5f$ shells
which contain almost localized electrons, and a broad conduction band of 
$s$, $p$ and $d$-electrons. In the periodic Anderson model, this
is simulated by an atomic-like level with strong on-site Coulomb
interaction which hybridizes with an uncorrelated conduction
band. Usually the model is further simplified by assuming 
both the atomic- and conduction states to be non-degenerate. The on-site Coulomb
interaction within the  $f$-states has to be considered as the largest
energy scale of the system since the electrons occupying these states
are less screened than the conduction electrons. In the PAM, the Coulomb 
interaction is further necessary to ensure that the $f$-levels are not
fully occupied, thus resembling the situation found in the Lanthanides
and Actinides.
Due to the incompletely filled $f$-levels, local
moments could be formed at every lattice site. It has been a
long-standing question whether these moments order magnetically or 
whether the local moments are screened by conduction band
electrons (Kondo screening)\cite{Don77}. In this paper we will focus on
those parameters where the $f$ electron density is non-integer
(intermediate valence regime). In this parameter regime, we find a
stable ferromagnetic phase and
investigate its properties.

Kondo screening has been the subject of extensive
investigations\cite{BFGS87,Jar95,TJF97,Noz98,HMS99,TJPF99,PBJ00a,VTJK00,PBJ00b,MN00b,hewson}.
Besides few exact statements\cite{TSU97,YS93}, several approximate or
numerical methods have been applied to
explore
the possibility of ferromagnetism in the PAM, e.\ g.\ Hartree-Fock
calculations\cite{LM78}, spectral density approach (SDA)\cite{MNRR98,MN99a},
slave-boson techniques\cite{MW93,DS97,DS98}
and dynamic 
mean-field theory (DMFT)\cite{TJF97,And99}. 
The first methods have severe limitations concerning the description of
the low-energy physics. The DMFT-based method promise an improvement at
this point. However, there are still only few results available on
ferromagnetism in the PAM within dynamical mean-field theory.

The antiferromagnetic phase of the PAM has been discussed
in more
detail\cite{MW93,Jar95,HC96,TJF97,DS97,DS98,And99}. It
seems to be widely accepted that
close to the symmetric parameter conditions, i.\ e.\ half-filling and
symmetric DOS, antiferromagnetism has to be
expected. This can also be concluded from the following: The PAM can be
mapped onto an effective Hubbard model\cite{MNRR98,MN99a}. In the
vicinity of the
symmetric point
the effective Hubbard model will also be close to half-filling. In that case
the Hubbard model is commonly expected to show
antiferromagnetic ordering\cite{And63}, therefore the PAM will also have
a tendency
towards antiferromagnetic ordering as was indeed found\cite{TJF97}.
Ferromagnetism is expected further away from this symmetric
point. A ferromagnetic phase was found for the PAM both in the Kondo
regime\cite{MW93,YS93,TJF97,DS98} and in the intermediate valence
region\cite{And99,MNRR98,MN99a}. Whereas in the former case,
the driving force towards ferromagnetism can be understood via an
effective Heisenberg-like coupling of the $f$-spins (RKKY
interaction)\cite{TJF97}, such a picture is not easily applied to the IV
regime due to the non-integer density of $f$-electrons and their
effective itineracy.

In this paper, we will investigate ferromagnetism in the PAM using
dynamical mean-field theory\cite{MV89,PJF95,GKKR96}. This theory is
based on the assumption of a $\vec{k}$-independent self-energy, which
becomes exact in the limit of infinite dimensions
($d=\infty$)\cite{MV89}. As pointed out in reference~\onlinecite{Geb91}, 
in this limit the lattice coherence and the exchange effects due to the
Pauli principle are preserved, contrary to other approaches based on
$\frac{1}{N}$ expansion ($N$ being the degeneracy of the model)\cite{Col83,RN83}.
Especially the exchange effects should be vital for
ferromagnetism, therefore the DMFT seems a method of choice.

In the next section, the DMFT together with the \textit{modified
  perturbation theory} (MPT) are introduced. The DMFT will lead to a
mapping of the PAM onto a single-impurity model, which then will be
solved by applying the MPT.
The results
concerning zero temperature as well as finite temperatures will be
presented and discussed in section~\ref{sec:Results-Discussion}.

\section{DMFT and the modified perturbation theory}
\label{sec:DMFT-modif-pert}
The periodic Anderson model is defined by its Hamiltonian
\begin{align}
  \label{hamiltonian}
    H =&\sum_{\vec{k},\sigma} \epsilon(\vec{k})s_{\vec{k}\sigma}^{\dagger}s_{\vec{k}\sigma} +
    \sum_{i,\sigma} e_f f_{i\sigma}^{\dagger}f_{i\sigma} +\\ 
    &V \sum_{i,\sigma} (f_{i\sigma}^{\dagger}s_{i\sigma} +
    s_{i\sigma}^{\dagger}f_{i\sigma} ) + \frac{1}{2} U \sum_{i,\sigma}
    n_{i\sigma}^{(f)}n_{i-\sigma}^{(f)}\nonumber
\end{align}
$s_{\vec{k}\sigma}$ ($f_{i\sigma}$) and $s_{\vec{k}\sigma}^{\dagger}$
($f_{i\sigma}^{\dagger}$) are the annihilation and creation operators
for an electron in a non-degenerate conduction band state (localized
$f$-state), and $n_{i\sigma}^{(f)}=f_{i\sigma}^{\dagger}f_{i\sigma}$
is the occupation number operator for the $f$-states. The dispersion $\epsilon(\vec{k})$
describes the propagation of free, i.\ e.\ unhybridized
conduction electrons, $e_f$ is the
position of the free $f$-level relative to the center of mass of the
conduction band density of states. The hybridization $V$ is taken as a
real, $\vec{k}$-independent constant, and finally $U$ is the Coulomb repulsion between two
$f$-electrons on the same lattice site.

The quantity of interest will be the $f$-electron Green function
\begin{equation}
  \label{eq:greenfunction}
    G_{ii\sigma}^{(f)}(E) = \langle\!\langle f_{i\sigma} ;
    f_{i\sigma}^{\dagger} \rangle\!\rangle
    = \sum_{\vec{k}} \frac{1}
    {E-(e_f-\mu)- \frac{V^2}{E-(\epsilon(\vec{k})-\mu)}
      -\Sigma_{\vec{k}\sigma}(E)}  
\end{equation}
To determine this function, we employ the dynamical mean-field theory
(DMFT)\cite{MV89,GKKR96,PJF95}. It was shown that in the limit $d\rightarrow\infty$, the
self-energy $\Sigma_{\vec{k}\sigma}(E)$ becomes purely local, i.\ e.\
$\vec{k}$-independent\cite{MV89,Mue89}. In this case, the self-energy of the PAM is
equivalent to the self-energy of a properly defined single-impurity
Anderson model (SIAM)\cite{Ohk91,PJF95,GKKR96}. The latter has to be defined
by the so-called self-consistency condition
\begin{equation}
  \label{eq:selfconsistency}
    \Delta_{\sigma}(E)= E-(e_f-\mu)-\Sigma_{\sigma}(E)
  -\left(G_{ii\sigma}^{(f)}(E)\right)^{-1} 
\end{equation}
instead of the usual definition
$\Delta=\sum_{\vec{k}}\frac{V^2}{E-(\epsilon(\vec{k})-\mu)}$ for the pure
SIAM.  Using
perturbation theory it has been shown in references~\onlinecite{SC89b,SC90a} that for
a three-dimensional system, the local approximation (equivalent to the
limit $d=\infty$) provides already for satisfactory results.

Now one is left with the problem to solve the SIAM defined by
equation~(\ref{eq:selfconsistency}). Here we use the \textit{modified
  perturbation theory} (MPT) which can be understood as an improvement
of the IPT scheme introduced in
references~\onlinecite{GKKR96,KK96}. This method was presented in more
detail elsewhere\cite{PWN97,MWPN99}, so we will restrict ourselves to a
short summary here.
Starting point is the following ansatz for the
self-energy\cite{MR82,MR86}:
\begin{equation}
  \label{eq:ansatz}
  \Sigma_{\sigma}(E)=U \langle n_{-\sigma}^{(f)}\rangle
  +\frac{\alpha_{\sigma} \Sigma_{\sigma}^{\rm (SOC)}(E)}
  {1-\beta_{\sigma} \Sigma_{\sigma}^{\rm (SOC)}(E)}
\end{equation}
$\alpha_{\sigma}$ and
$\beta_{\sigma}$ are introduced as parameters to be determined
later. $\Sigma_{\sigma}^{\rm (SOC)}(E)$ is the
second-order contribution to perturbation theory around the Hartree-Fock 
solution\cite{Yam75,ZH83,SC90a}. The Hartree-Fock
solution introduces another free parameter, namely the chemical
potential within this calculation: $\tilde{\mu}$. \textit{A
  priori} it is 
not clear whether this should be equal to the chemical potential in the
full (DMFT-MPT) calculation, or whether e.\ g.\ it should be determined such that
the electron density on the impurity site of the SIAM is equal both
for the Hartree-Fock and DMFT-MPT calculation. In
reference~\onlinecite{KK96} and other papers\cite{TS99b,VTJK00}
yet another condition was used to determine $\tilde{\mu}$. There, the
Luttinger theorem\cite{LW60} or equivalently the Friedel sum
rule\cite{Fri56,Lan66} was forced to hold. Since these theorems are applicable
only for $T=0$, this limits the calculations to zero temperature. In
order to access finite temperatures, we used the condition of
identical electron densities for the Hartree-Fock and 
the full calculation ($n_{\sigma}^{(f,{\rm HF})}=n_{\sigma}^{(f)}$).
With this choice, the Friedel sum 
rule is still fulfilled in a large parameter region as could be shown
for the pure SIAM\cite{MWPN99}. A more detailed analysis of the
different possibilities to determine $\tilde{\mu}$ is found in
reference~\onlinecite{PWN97}. 

Next, the remaining parameters $\alpha_{\sigma}$ and $\beta_{\sigma}$
have to be determined. Instead of using the ``atomic'' limit of $V=0$ as 
was done e.\ g.\ in references~\onlinecite{KK96,VTJK00,Sas99pre2} we
make use of the moments of the spectral density
\begin{gather}
  \label{eq:moments}
  M_{\sigma}^{(n)}=
  \int \! dE \, E^n A^{(f)}_{\sigma}(E) =\langle  [
  \underbrace{[...[f_{\sigma},H]_-,...,H]_-}_{ \text{$n$-fold
      commutator}} , f_{\sigma}^{\dagger} ]_+\rangle
  \\ 
  A^{(f)}_{\sigma}(E)=-\frac{1}{\pi}\Im G^{(f)}_{ii\sigma}(E+i0^+)\nonumber
\end{gather}
where $[ A;B]_-$ ($[ A;B]_+$) denotes the commutator (anticommutator) of 
the operators $A$ and $B$ and $\Im x$ denotes the imaginary part of $x$.
As indicated, these can be calculated on two different ways, therefore
conditions (sum rules) can be derived. To determine $\alpha_{\sigma}$
and $\beta_{\sigma}$, the first four moments $n\in\{0,\dots3\}$ have to
be used since the $n=0$- and $n=1$-moments are fulfilled for any
$\alpha_{\sigma}$ and $\beta_{\sigma}$. In the $n=3$-moment a
higher-order correlation function that we will call \textit{bandshift}
$B_{\sigma}$ is introduced:
\begin{eqnarray}
  \label{eq:band-shift}
  \lefteqn{\langle n^{(f)}_{\sigma}\rangle(1-\langle n^{(f)}_{\sigma}\rangle)
    (B_{\sigma}-e_f)=}\nonumber\\ &&
    =\sum_k V_{k {\rm d}}\langle
    s_{k\sigma}^\dag f_{\sigma}(2n^{(f)}_{-\sigma}-1)\rangle
  \\ & &
  =
  -\frac{1}{\pi}\Im \int \! dE \, f_-(E)\Delta_{\sigma}(E)
  \left(\frac{2}{U}\Sigma_{\sigma}(E)-1\right)G^{(f)}_{ii\sigma}(E)
  \nonumber
\end{eqnarray}
with the Fermi function $f_-(E)=(\exp(\beta E)+1)^{-1}$.

Now, the solution of the SIAM by the MPT has to be integrated into the
DMFT-self-consistency loop\cite{GKKR96}: Starting with a guessed value
of $\Sigma_{\sigma}(E)$, equations (\ref{eq:greenfunction}) and
(\ref{eq:selfconsistency}) are evaluated, and then the new MPT
self-energy (\ref{eq:ansatz}) is calculated for the appropriate
SIAM. This procedure is iterated until a self-consistent solution is found.
Within this formalism, the spontaneous symmetry-breaking necessary for
ferromagnetic solutions can be introduced via spin-asymmetric starting
values for the self-consistency cycle. In general, it is not necessary to
introduce a symmetry-breaking zero-field. 

At this point, let us comment on the quality of our approximative
method: Being based on perturbation theory, the MPT is expected to give
reliable results for small interaction strengths. For the symmetric
SIAM, it was even shown\cite{ZH83} that the perturbative expansion is
essentially equivalent to the exact Bethe-ansatz
solution\cite{And80,TW83} and therefore also valid in the strong coupling
regime. For asymmetric parameters, we improve on the perturbation theory
by enforcing the correct high-energy expansion for the self-energy and
equivalently the Green functions. This is archieved by fulfilling the
first four
sum rules defined by the spectal
moments~(\ref{eq:moments}) which leads to the correct determination of
the spin-dependent positions and weights of the charge
excitations up the order $\frac{1}{U}$ in accordance with
reference \onlinecite{HL67}\cite{PHWN98}.
From this, we derive our proposition that the MPT
can also be reasonably applied to a PAM in the intermediate-to-strong
coupling regime. Although for the high-energy (high-temperature)
behaviour this follows from the discussion above,
the quality  of the special low-energy properties of the PAM within the
MPT is not
known \textit{a priori}.
To estimate
the significance of the results for the intermediate coupling
strengths, comparison with
exact or numerically exact methods is necessary.
For example, one can 
apply the MPT to a SIAM without the context of the DMFT and compare with
the exactly known properties of that model\cite{hewson}. This comparison
was done in reference~\onlinecite{MWPN99} and the results can be
summarized as follows: 
The charge excitations are at the
proper positions for symmetric as well as asymmetic
parameters. Concerning the low-energy behaviour, there is a qualitative
agreement but quantitative discrepancies: The Friedel sum rule is
fulfilled in a large parameter space off the symmetric point, 
but the
Kondo temperature does not follow an exponential law. A power law is
found instead. To summarize, the MPT does qualitatively include
essential parts of the Kondo physics, but is prone to deviations
concerning energy (temperature) scales. Further tests of the MPT can be
done by 
comparing results for the PAM with different methods based on DMFT, as
e.\ g.\ the numerically exact quantum Monte Carlo
(QMC)\cite{Jar95,TJF97}, exact diagonalization (ED)\cite{GKKR96} or
numerical renormalization group theory (NRG)\cite{PBJ00b}. Comparisons
of this kind for the paramagnetic PAM have been published in
reference~\onlinecite{VTJK00} for QMC and reference~\onlinecite{Sas97}
for ED calculations. Also the results of
reference~\onlinecite{MN00b} can easily be compared with above-mentioned
references. Again, the qualitative features
compare well, however temperature scales concerning low-energy
properties cannot be reproduced quantitatively. Also, the
$V$-dependence seems to be overestimated by the MPT when compared to
NRG\cite{Pruschkeprivate}. 
We do believe that these shortcomings of the MPT do not inhibit
the analysis of the ferromagnetic properties of the PAM. In the
discussion of our results, we will substanciate this claim.

In this paper,
we use the following system parameters: The conduction band is
described by a semi-elliptic density of states of unit width centered
at $E=0$, thus defining the energy scale. The position of the
$f$-level, $e_f$ will be given relatively to the center of gravity of the
conduction band. The hybridization is taken to be constant. Unless
otherwise noted, we chose $V=0.2$. The Coulomb interaction strength
$U$ is set to $U=4$ for most calculations, thus representing the
largest energy in the system. The total number of
electrons $n^{\text{(tot)}}$, $e_f$ and the temperature $T$ will be varied.
The latter will be given in $K/eV$ within the energy scale
defined by the width of the conduction band.

\section{Results and Discussion} 
\label{sec:Results-Discussion}
Before discussing the results concerning ferromagnetism in the PAM, let
us recall some results for the paramagnetic case as they will be
important for the following. The density of states (DOS) consists of
the two charge excitation peaks approximately at
$e_f$
and $e_f+U$. In addition, a sharp resonance close to the chemical
potential is induced by the correlations, the Kondo resonance. Due to
lattice coherence, the Kondo resonance is split by the coherence
gap in case of symmetric parameters. 
Another result for the paramagnetic PAM found within the
DMFT-MPT scheme is the existence of local Kondo singlets for large
hybridization strengths\cite{MN00b}. More detailed reviews on the
paramagnetic PAM can be
found in references
\onlinecite{hewson,Czy86,SC89a,SC90b,Geb91,Jar95,HMS99,MN00b}.

\begin{figure}[b]
  \begin{center}
    \epsfig{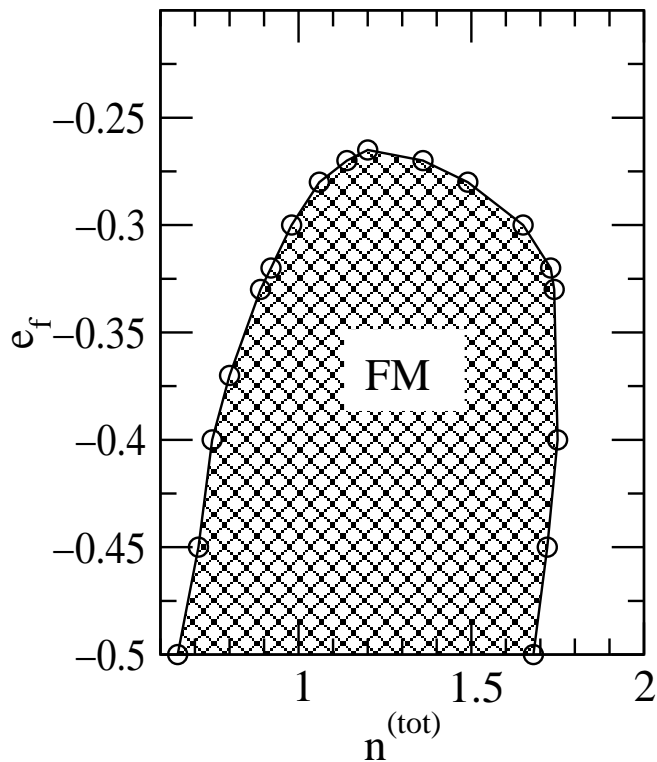}
    \caption{ $e_f$ vs.\ $n^{\text{(tot)}}$ phase diagram for $T=0$, $U=4$,
      $V=0.2$}
    \label{fig:phdia}
  \end{center}
\end{figure}

In figure~\ref{fig:phdia}, the region of ferromagnetic order is plotted
in the $e_f$ vs.\ $n^{\text{(tot)}}$ phase diagram.
We have restricted our evaluation  to the
so-called intermediate valence region (IV region) by positioning
the
$f$-level within the conduction band which extends from 
$-0.5$ to $0.5$. This results in a
$f$-occupancy $n^{(f)}$ smaller than unity. Previous investigations have
shown that ferromagnetism can exist in this parameter
regime\cite{RU85,And99,MNRR98}.
The opposite case is the
so-called Kondo regime, which is obtained by taking $e_f$ well below the
lower conduction band edge, and $U$ sufficiently large so that $e_f+U$
is clearly above the conduction band. This leads to a nearly
half-filled $f$-level. 
As discussed in the introduction, a half-filled $f$-level should
introduce a tendency towards antiferromagnetism which we have not yet
investigated. 
In the following we will therefore concentrate on
ferromagnetism in the 
intermediate valence region, but we will also be noting some trends on how the
ferromagnetic phase would continue into the Kondo
regime.
As can be seen in figure~\ref{fig:phdia}, the region of
ferromagnetic order is quite large, determined by an upper bound for
$e_f$ and upper and lower bounds for $n^{\text{(tot)}}$. As will be shown
below, the ferromagnetic region continues into the Kondo regime.

\begin{figure}[b]
  \begin{center}
    \epsfig{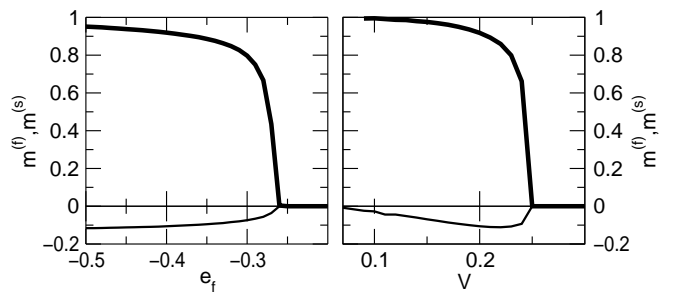}
    \caption{$f$- ($s$) -magnetization as
      function of the hybridization strength $V$ and $f$-level position
      $e_f$ for $T=0$, $n^{\text{(tot)}}=1.3$ and $U=4$ plotted in thick (thin)
        lines (left panel: $V=0.2$, right panel: $e_f=-0.4$).} 
    \label{fig:m_vef}
  \end{center}
\end{figure}

In the left panel of figure~\ref{fig:m_vef}, the $f$- ($s$-)
magnetization,
$m^{(f,s)}= \frac{n^{(f,s)}_{\uparrow}-n^{(f,s)}_{\downarrow}} {n^{(f,s)}_{\uparrow}+n^{(f,s)}_{\downarrow}}$ 
is plotted as    
function of $f$-level position 
$e_f$. The parameters are as in figure~\ref{fig:phdia}, the band
occupation is given by $n^{\text{(tot)}}=1.3$. Ferromagnetic order breaks down for $e_f$
approaching the center of gravity of the conduction band as it is
commonly expected\cite{RU85}. Within our numerical accuracy, this
quantum phase transition is of second order.

In the other limit, i.\ e.\ $e_f$ below the lower edge of the conduction 
band ($e_f<-0.5$), the ferromagnetic solution remains stable. We
confirmed that down to
$e_f=-1.2$ the magnetization behaves as one would extrapolate from figure
\ref{fig:m_vef}. So, the ferromagnetic phase plotted in
figure~\ref{fig:phdia} continues down into the Kondo regime and connects
to the ferromagnetic
phase found in reference~\onlinecite{TJF97}. There, the
authors found a stable ferromagnetic phase for $n^{\text{(tot)}}<1.6$.

The right picture of figure~\ref{fig:m_vef} shows the magnetization as
function of the hybridization strength for $e_f=-0.4$, the remaining
parameters as in the left panel. The 
ferromagnetic order is destroyed by large hybridizations. This can be
understood in terms of local singlet formation. If the conduction
electrons screen 
the magnetic moment of the localized electrons, as discussed in detail
in reference~\onlinecite{MN00b}, magnetic ordering will not be
possible any more. 
Comparison with NRG results\cite{PBJ00b,Pruschkeprivate} seem to
suggest that the $V$-dependence of certain ``low-energy'' quantities is
over-estimated by the MPT, as e.\ g.\ the size of the coherence gap. This
implies the possibility that the critical $V$ is
underestimated since the formation of local Kondo singlets is due to the
low-energy physics of the PAM.
For
$V\rightarrow0$ the ferromagnetic phase is stable, for numerical reasons 
no calculations for $V\lesssim 0.03$ have been performed.
For $V=0$, the $f$-level and the conduction band decouple completely.
The $f$-level basically corresponds to the zero-bandwidth Hubbard
model\cite{NolBd7} in which ferromagnetism is not stable.
In figure~\ref{fig:m_vef}, the thin lines represent the conduction band
polarization $m^{(s)}$. In general, it is found to be
antiparallel to $m^{(f)}$. However, as will be shown below, a parallel
alignment of $m^{(f)}$ and $m^{(s)}$ is also possible for certain parameters.

\begin{figure}[tb]
  \begin{center}
    \epsfig{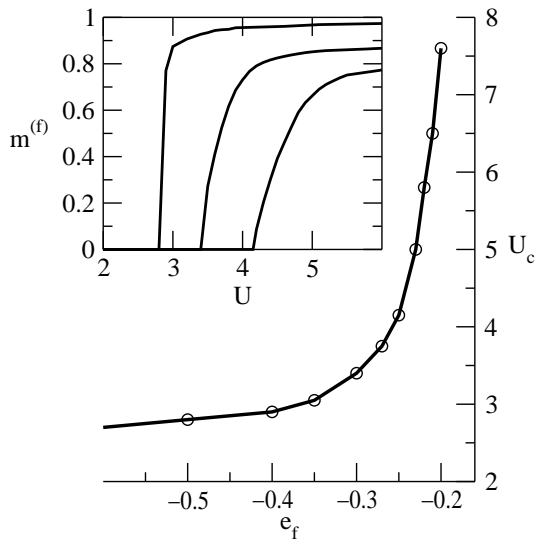}
    \caption{Critical interaction strength $U_{\rm c}$ as function of
      $e_f$ for $T=0$, $V=0.2$ and $n^{\text{(tot)}}=1.3$.
      The inset shows the respective $f$-magnetization as function of $U$ for
      $e_f=-0.5,-0.3,-0.25$ (from top to bottom).} 
    \label{fig:Uc}
  \end{center}
\end{figure}

In figure~\ref{fig:Uc}, the $U$-dependence is examined. The inset shows
that ferromagnetism sets in at a critical $U_{\rm c}$ and saturates
quickly afterwards.
The absolute value of $U_{\rm c}$ varies strongly
with $e_f$: On 
shifting $e_f$ below the conduction band into the Kondo regime, $U_{\rm
  c}$ seems to saturate at a value of $U_{\rm c}^{\text{(KR)}}\approx
2.5$. 
However, there is no true saturation, because when $e_f$ is low enough
so that the upper charge excitation is close to or even below the
chemical potential, the physics of the system 
will change again.
On increasing $e_f$, $U_{\rm c}$ increases as well and exhibits a
sharp upturn at $e_f\rightarrow -0.2$. 
In reference~\onlinecite{TJF97}, a ferromagnetic phase was found in the
Kondo regime of the PAM for interaction strengths smaller than our 
$U_{\rm c}^{\text{(KR)}}$. As for the hybridization, the MPT seems to
misjudge the energy scale. We will argue below that the Kondo scale is
decisive for ferromagnetism in the Kondo regime\cite{TJF97}, but not
in the intermediate valence regime.

\begin{figure}[htb]
  \begin{center}
    \epsfig{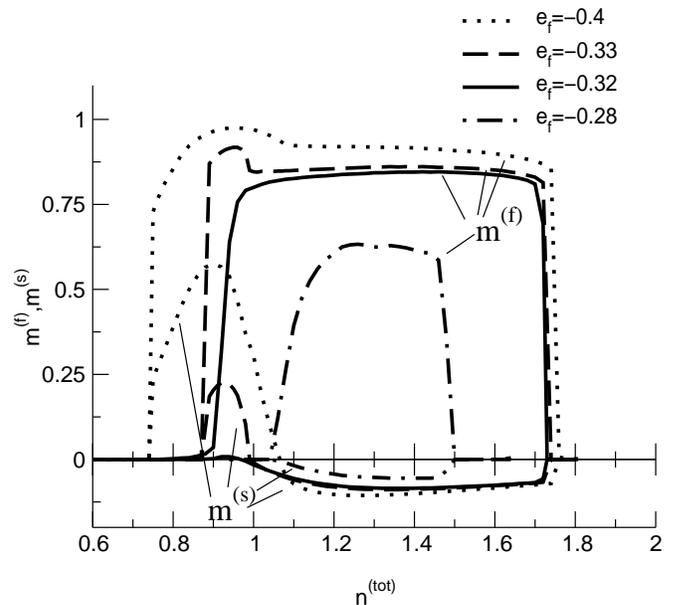}
    \caption{$f$- and $s$-magnetization as function of electron density
      for various $e_f$ and $U=4$, $V=0.2$ and $T=0$.}
    \label{fig:m_n}
  \end{center}
\end{figure}

In figure \ref{fig:m_n}, 
both the $f$- and $s$-magnetization are plotted as function of the
electron density for various values of $e_f$ below half-filling
($n^{\text{(tot)}}<2.0$). As already seen in the phase diagram, the system is
ferromagnetic in a range of values of $n^{\text{(tot)}}$ below half-filling
$n^{\text{(tot)}}=2.0$. Both the lower and upper boundary of the ferromagnetic
  region depend on $e_f$.
The lower boundary is in fact determined by the
number of $f$-electrons $n^{(f)}$, which itself depends on $e_f$ and
$n^{\text{(tot)}}$. The phase transition occurs for each $e_f$ at that
specific electron density, where $n^{(f)}$ drops below $0.56\pm0.02$.
Approaching the half-filled system, no ferromagnetism is found. This is
similar to the 
findings in reference \onlinecite{TJF97}. However as discussed above, 
antiferromagnetic order is to be
expected\cite{MW93,Jar95,HC96,TJF97,DS98} in this case.

For $e_f<-0.32$ the ferromagnetic phase can be separated into two
regions of different properties. In the low-density regime, the
$f$-polarization tends to saturation, the $s$-polarization is
positive. On the contrary, in the high-density region ($1.1\lesssim
n^{\text{(tot)}}\lesssim 1.7$), the $f$-polarization is almost constant, below saturation,
and the conduction band polarization is of opposite sign
(antiferromagnetic coupling between $f$- and $s$-band). For more clarity, 
we will call the first case parallel, the latter case antiparallel
coupling of $f$- and $s$-electrons. This latter case of antiparallel
coupling is generally expected for the PAM due to its close relation to
the Kondo lattice model via the Schrieffer-Wolff
transformation\cite{SW66,hewson}. This transformation maps the PAM onto 
a model with an antiferromagnetic interband-spin exchange.
One condition for the Schrieffer-Wolff
transformation is that the $f$-level should be half-filled, which is not
met here. However, for large total occupation number, $n^{(f)}$ is
closer to unity than for small $n^{\text{(tot)}}$. Therefore the finding of
parallel coupling for low, and antiparallel coupling for high electron
densities is no contradiction to the Schrieffer-Wolff
transformation.

\begin{figure}[htb]
  \begin{center}
    \epsfig{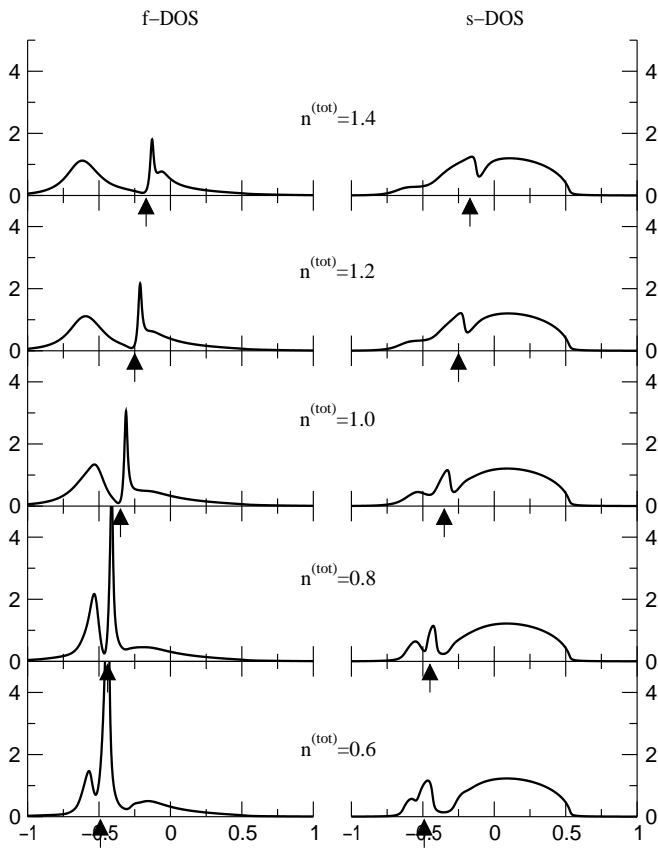}
    \caption{Paramagnetic $f$- and $s$-densities of states for
      $e_f=-0.4$, $V=0.2$,
      $U=4$ and $T=0$. The different panels correspond to
      different electron densities $n^{\text{(tot)}}$}
    \label{fig:qdos_para}
  \end{center}
\end{figure}

\begin{figure}[htb]
  \begin{center}
    \epsfig{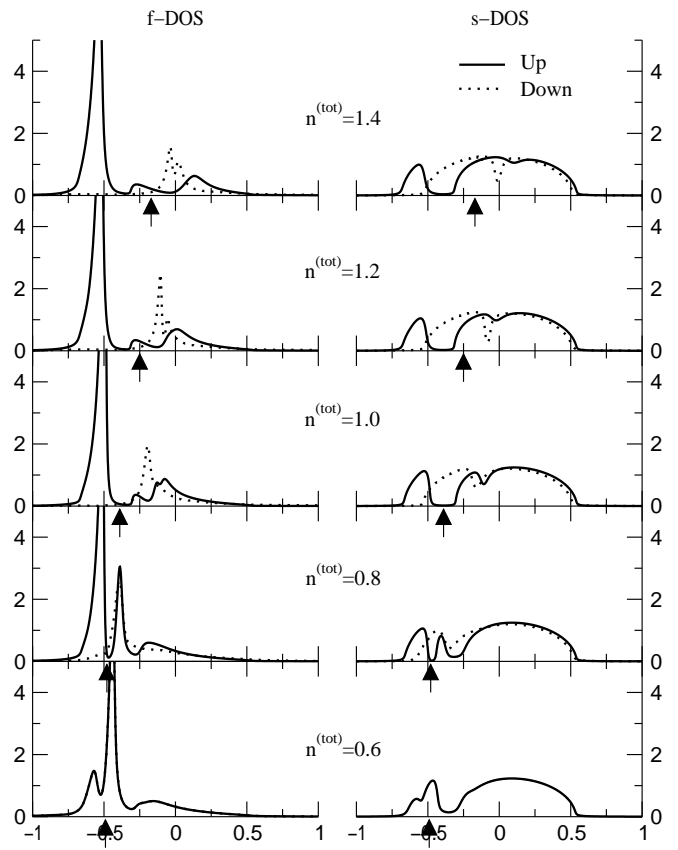}
    \caption{same as figure~\ref{fig:qdos_para}, but showing the (stable)
      ferromagnetic solution. The parameters are as in
      figure~\ref{fig:qdos_para}, thus corresponding to the dotted line of
      figure~\ref{fig:m_n}. Solid line: spin $\uparrow$; dotted
      line: spin $\downarrow$.}
    \label{fig:qdos_ferro}
  \end{center}
\end{figure}

In figures~\ref{fig:qdos_para} and~\ref{fig:qdos_ferro}, the densities
of states are plotted for various band
occupations with $e_f=-0.4$, $V=0.2$, $U=4$ and zero temperature. The
respective
left panel corresponds to a projection onto $f$-, the right
panel onto $s$-states.
The upper charge excitation, located
approximately at $E\approx e_f+U=3.6$ is not visible in these
figures. The arrows indicate the respective position of the chemical
potential. To get more  
insight into the complicated structure of the DOS, we start by
explaining figure~\ref{fig:qdos_para}, where we have forced the
system to be paramagnetic. In figure~\ref{fig:qdos_ferro}, the stable
ferromagnetic DOS are shown and will be explained below.

Starting with the $f$-DOS in figure~\ref{fig:qdos_para}, one observes
three structures: the first is
the lower charge excitation located roughly at the lower band edge. With 
increasing electron density, this peak shifts to lower
energies. 
The next structure is 
the Kondo resonance which is in the vicinity of the chemical
potential $\mu$. Due to the asymmetry of the parameters, i.\ e.\ the two
charge excitation peaks are not symmetric around the chemical potential
and the band occupation is well below $n^{\text{(tot)}}=2$, the Kondo
resonance is not exactly at $E=\mu$ but slightly above. This is in
agreement with previous  findings\cite{VTJK00,MN00b} and was also observed in
the NRG method\cite{PBJ00b} and the single-impurity 
Anderson model\cite{MWPN99,hewson}. With increased electron density, the
Kondo resonance moves towards higher energies, this is clearly related
to the according shift of the chemical potential $\mu$. 
It is reasonable to expect the Kondo resonance to vanish for decreasing
$n^{(f)}$. Within the MPT, the Kondo resonance remains stable for all
parameters discussed in this paper. Only for much reduced $n^{(f)}$,
where $e_f$ is located well above $\mu$, the Kondo resonance disappears.
As already
mentioned, there is the ``coherence gap'' in the center of the Kondo
resonance for 
symmetric parameters. Due to the asymmetry of the parameters in
figure~\ref{fig:qdos_para}, this gap is closed. Only for the
$n^{\text{(tot)}}=1.4$ case (upper panel of figure~\ref{fig:qdos_para}),
there is a slight dip in the $f$-DOS visible. This complies with
the findings of reference~\onlinecite{PBJ00b}.
The third feature 
in the $f$-DOS is a broad structure representing states induced by the
hybridization with the conduction band. A fourth feature is of course
also present in the DOS, although not plotted in
figures~\ref{fig:qdos_para} and \ref{fig:qdos_ferro}: the upper charge
excitation which is located roughly at $e_f+U=3.6$ and has a FWHM (full
width at half maximum) of $0.05$.

All these structures have their
counterpart in the conduction-band DOS shown in the right panel of
figure~\ref{fig:qdos_para}. The charge excitation induces via the
hybridization states in the $s$-DOS. Where the Kondo resonance is
visible in the $f$-DOS, the $s$-DOS shows also some anomalies. The original
conduction band can also be clearly recognized.

Let us now discuss how the DOS are modified in the ferromagnetic phase. In
figure~\ref{fig:qdos_ferro} the (stable) ferromagnetic DOS are plotted
for the same parameters as in figure~\ref{fig:qdos_para}.
The solid lines represent the
spin-$\uparrow$, the dotted lines the spin-$\downarrow$ DOS. For the
lowest panel ($n^{\text{(tot)}}=0.6$), the system is paramagnetic. For the
ferromagnetic solutions, the following picture emerges: Firstly, the
lower charge excitation is fully polarized, i.\ e.\ it consists purely
of spin-$\uparrow$ states. Correspondingly, the upper charge excitation,
which is not visible, exists only in the spin-$\downarrow$ DOS.
Some interesting changes occur to the Kondo resonance: Whereas in the
spin-$\uparrow$ channel it disappears almost completely, there is still
a peaked structure visible in the spin-$\downarrow$ channel. 
Contrary to what one would expect for the Kondo resonance, its shift
towards higher energies on increasing $n^{\text{(tot)}}$ is stronger
than the according shift of $\mu$. It is therefore not located at $\mu$
any more.
Although this structure
clearly develops from the Kondo resonance, 
its physical interpretation
(Kondo screening) cannot so easily be transfered.
Finally there are some substantial changes in the $s$-DOS as
well:
Due to the full polarization of the lower charge excitation, the
hybridization-induced $s$-states in the same energy range are also
purely of majority spin. The feature that
corresponds to the Kondo resonance is, for similar reasons, only visible in the
spin-$\downarrow$-$s$-DOS. 
Finally, in the spin-$\uparrow$ DOS, a new
gap appears which separates the lower charge excitation from the remainder of 
the bare band. Due to this new gap, the system becomes a semi-metal
around quarter-filling ($n^{\text{(tot)}}=1.0$) meaning that
the
spin-$\uparrow$-DOS vanishes at $\mu$, so only
spin-$\downarrow$-electrons contribute to electrical current.
The new gap, which we call
hybridization gap is
not equivalent to the previously 
discussed coherence gap although the latter is of course also induced by
the hybridization. We will explain the differences between these gaps below.
This gap is also present in the $f$-DOS, but not visible in the plotted
figures because of an artificial broadening of the lower charge
excitation which is necessary for numerical reasons.

From these observations the change of sign of the conduction
band polarization can be understood: In the case of low electron
density, the chemical potential lies within the lower charge
excitation. As explained above, due to the hybridization there are only
spin-$\uparrow$ conduction band states available in the occupied region
of the DOS, the conduction band polarization is necessarily parallel to 
the $f$-polarization. But on the other hand, with increasing electron
density, $\mu$ is increasing as well. The chemical potential now has to
cross the hybridization gap which exists only in the spin-$\uparrow$
DOS. As a result, there are more occupied spin-$\downarrow$ than spin-$\uparrow$
states and the conduction band polarization is antiparallel to
the $f$-magnetization.

\begin{figure}[htb]
  \begin{center}
    \epsfig{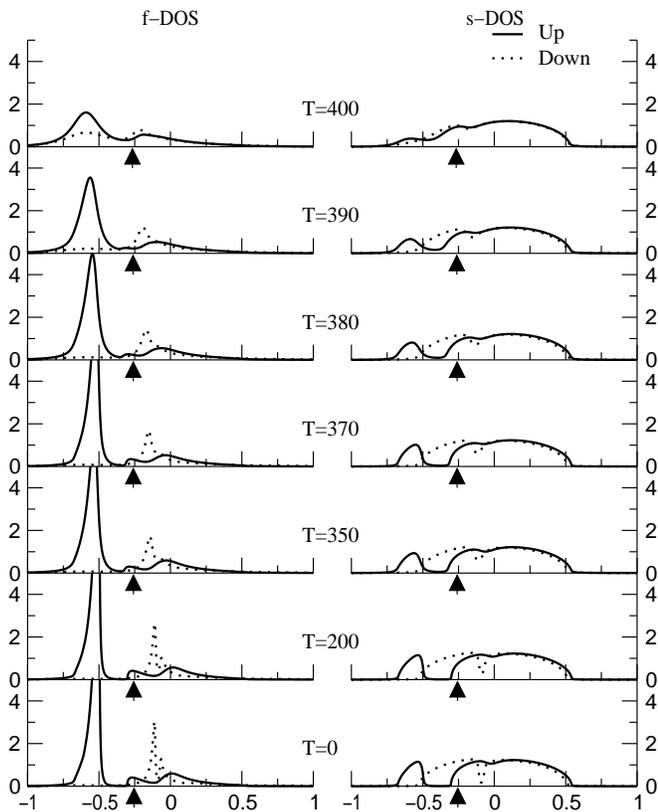}
    \caption{Temperature dependence of the DOS for $n^{\text{(tot)}}=1.2$. The
      remaining parameters are as in figure~\ref{fig:qdos_ferro}.}
    \label{fig:qdos_ferro_temp}
  \end{center}
\end{figure}

Let us now discuss the difference between the well-known coherence gap
and the above-described hybridization gap. The coherence gap is found
for a symmetric PAM with on-site hybridization\cite{Jar95}. Contrary to
the single-impurity case, a gap develops in the center of the
Kondo resonance located at $E=\mu$. Shifting away from symmetric
parameters, this coherence gap gets suppressed, although a dip within
the Kondo resonance is still visible for a large parameter
region\cite{VTJK00,MN00b,PBJ00b}. So why is the new gap observed in
figure \ref{fig:qdos_ferro} not this coherence gap? This question is
answered by the fact that the new gap does not shift with $\mu$ on
increasing the electron density. 
Although there is no proper Kondo
resonance in the case of ferromagnetic order, 
those structures that clearly develop from it do shift with increasing
$n^{\text{(tot)}}$. Within this structure, even a dip which is the relic
of the coherence gap is visible for $n^{\text{(tot)}}=1.4$. The position
of the new
gap is neither related to $\mu$ nor the relics of the Kondo resonance,
it is rather located at $e_f$.
And what is the origin of the new gap?
Since its position is determined by the value of $e_f$ and its width scales
with $V$, it reminds of the gap found e.\ g.\ within the
SDA\cite{MNRR98} or alloy analogy\cite{Czy86,RMSRN00pre}, but also for
the $U=0$ 
case of the PAM which is exactly
diagonalizable\cite{hewson}. This indicates that this gap is due to
level-repulsion between the lower $f$-peak and the conduction band, as
the coherence gap can be interpreted as level-repulsion between a virtual
$f$-level at the fermi energy and the conduction band.
To distinguish between these gaps, we call the former ``hybridization''
and the latter ``coherence'' gap.

At first sight, it surprises that this gap was not found in any
paramagnetic DMFT-based
calculation\cite{TJF98,PBJ00a,VTJK00,PBJ00b,MN00b}. 
This is in our opinion simply due to damping effects caused by
the correlation, i.\ e.\ a finite imaginary part of the self-energy.
In case of saturated ferromagnetism, $\Im \Sigma_{\uparrow} (E)$
vanishes and the corresponding spin-$\uparrow$ DOS should resemble the
$U=0$ DOS including the hybridization gap at $e_f$. In the
spin-$\downarrow$- or also the paramagnetic DOS, all 
structures are broadened due to the finite damping. This effectively
closes the hybridization gap for these DOS.
This proposal can be supported by investigating the temperature-dependence of the 
DOS. On increasing the temperature, the magnetization will decrease. The 
system is pushed away from saturation, damping effects occur also for
spin-$\uparrow$ electrons. In
figure~\ref{fig:qdos_ferro_temp}, the DOS for $n^{\text{(tot)}}=1.2$ are
plotted for various temperatures (the remaining parameters as in
figure~\ref{fig:qdos_ferro}). For these parameters, the Curie
temperature is $T_c\approx 408$. One can see how the gap closes on
approaching $T_c$. 
More complicated is the case of $n^{\text{(tot)}}=0.8$ in
figure~\ref{fig:qdos_ferro}. On the one hand, ferromagnetic saturation
is not reached (cf.\ figure~\ref{fig:m_n}) and the hybridization gap is
partly closed by damping. And on the other hand, the chemical potential
is located in the same energy range, further complicating the situation.

\begin{figure}[htb]
  \begin{center}
    \epsfig{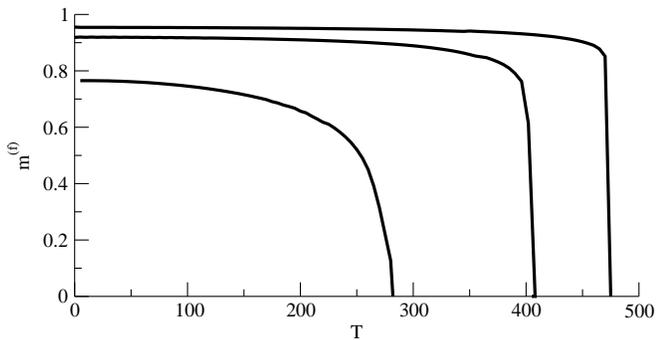}
    \caption{Magnetization curves for $U=4$, $V=0.2$, $n^{\text{(tot)}}=1.2$
      and various $e_f \in\{-0.5,-0.4,-0.3\}$ (from top to bottom)}
    \label{fig:m_t}
  \end{center}
\end{figure}
\begin{figure}[htbp]
  \begin{center}
    \epsfig{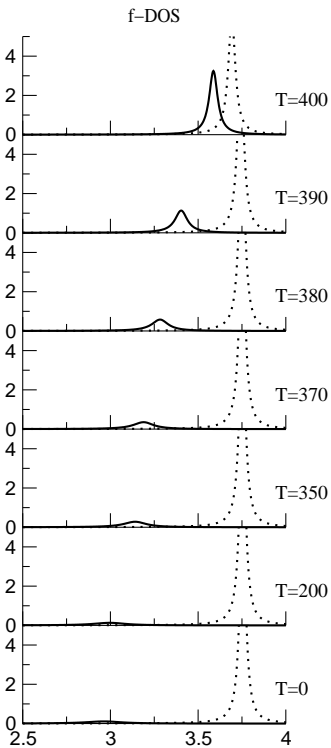}
    \caption{Upper charge excitation peak of the $f$-DOS for the
      same parameters as in figure~\ref{fig:qdos_ferro_temp}}
    \label{fig:qd_t_upper}
  \end{center}
\end{figure}

In the following we will discuss the temperature dependence in more detail.
Figure \ref{fig:m_t} shows magnetization curves for
$n^{\text{(tot)}}=1.2$ and three different values of
$e_f\in\{-0.5,-0.4,-0.3\}$
(from top to bottom). The lowest line ($e_f=-0.3$) shows 
a clear second-order phase transition. The middle case corresponds to 
the data of figure \ref{fig:qdos_ferro_temp}.
The phase transition at $T_c\approx 408$ is
within numerical accuracy of second order. 
The demagnetization process is dominated by a transfer of spectral
weight from the minority- to the majority-spin spectrum of the lower
charge excitation peak as can be seen in
figure~\ref{fig:qdos_ferro_temp}. However, for the upper charge
excitation, plotted in figure~\ref{fig:qd_t_upper} for the same
parameters as figure~\ref{fig:qdos_ferro_temp}, not only the reverse
transfer of spectral weight occurs but also a spin-dependent
bandshift can be observed.
When decreasing $e_f$ towards the lower band edge of the conduction
band, the phase transitions become sharper, and finally,
first order phase transitions are found for $e_f \lesssim -0.5$ within
numerical accuracy. This is similar to findings for the Hubbard
model\cite{PHWN98}, where phase transitions were found to be of second
order for lower, and of first order for higher electron densities. In
the PAM, the density of correlated electrons, $n^{(f)}$, increases on
lowering $e_f$, therefore the behaviour is analogous to the findings for
the Hubbard model. As discussed in reference~\onlinecite{PHWN98}, we
cannot exclude that this is an artifact of our method.
\begin{figure}[htb]
  \begin{center}
    \epsfig{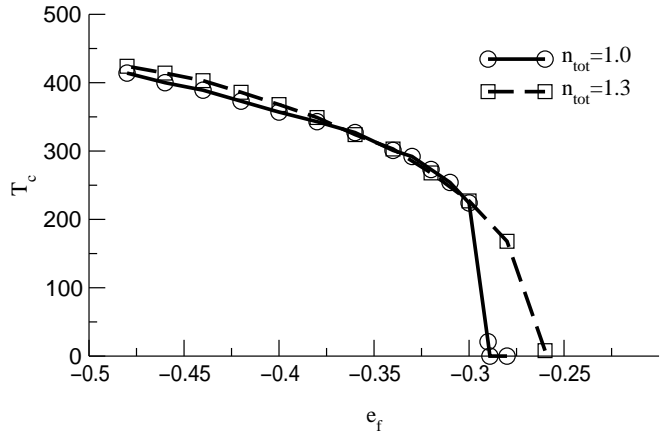}
    \caption{Curie temperatures as function of $e_f$ for $U=4$ and
      $V=0.2$. The electron density is as indicated.}
    \label{fig:tc_ef}
  \end{center}
\end{figure}
In figure~\ref{fig:tc_ef}, the Curie temperature is plotted as function of 
the $f$-level position. The trends are very similar to the
$T=0$-magnetization as seen in figure~\ref{fig:m_vef}. For $e_f$
approaching the upper critical value, both $T_c$ and $m(T=0)$ vanish. In
the opposite direction, both quantities keep increasing, however, with
diminishing slope. 
The Curie temperature as function of band occupation is shown in
figure~\ref{fig:tc_n} for $e_f=-0.4$ and $e_f=-0.3$. 

As already mentioned in the introduction, the magnetism in the Kondo
limit of the PAM can be understood in terms of an effective RKKY
interaction\cite{TJF97,NRM97} between the $f$-spins.
However, in the intermediate-valence regime under investigation here,
the effective RKKY exchange should be suppressed due to the non-integer
$f$-electron filling and the effective itineracy of the $f$-electrons as
indicated by the broader $f$-peaks in the DOS. We believe that 
instead of the
interband-RKKY exchange, a ``narrow-band'' mechanism is the main driving
force towards ferromagnetism in the IV regime.
The effective itineracy of the
$f$-electrons due to the hybridization in combination with the strong
on-site Coulomb interaction leads to the strong similarities to the 
well-known Hubbard model\cite{Hub63,Gut63,Kan63}. Recently
the existence of ferromagnetism was shown within the Hubbard
model\cite{Ulm98,OPK97,PHWN98}. One important condition
for it
is the shape of the free, i.\ e.\ $U=0$ DOS entering the model. A highly 
asymmetric DOS with
a divergence or sharp peak close to the band
edge enhances the possibility of ferromagnetic order\cite{Vea97}.
Turning back to the periodic Anderson model we note that this condition
favoring ferromagnetic order is excellently met for the
hybridization-broadened $f$-level in combination with the
hybridization-induced $f$-states within the conduction band.
Our proposal, that ferromagnetism in the IV regime of the PAM originates
from an intraband mechanism is
supported by several findings: 

\noindent i) Whereas in the Kondo regime, small values of $U$ can
already lead to a 
ferromagnetic phase, in the IV regime, significantly larger values of
$U$ are necessary.

\noindent ii) The particular low-energy physics, that are decisive for
ferromagnetism in the Kondo regime, have no
significant influence on the magnetism
in the IV regime.

\noindent iii) The critical $n^{\text{(tot)}}$, where ferromagnetism
breaks down, is in fact determined by a critical number of correlated
$f$-electrons, which is of similar magnitude as for the Hubbard model.

\noindent iv) The polarization of the conduction band seems to have
almost no influence on the magnetic properties in the IV regime.

Let us discuss these points in more detail:
The first point follows directly from figure~\ref{fig:Uc} and
reference~\onlinecite{TJF97}.
In the Kondo regime, where the low-energy physics determine the magnetic
properties\cite{TJF97}, the MPT probably overestimates the 
absolute value of $U_{\rm c}$. However, the sharp increase of $U_{\rm
  c}$ for increasing $e_f$ indicates the importance of a strong-coupling
mechanism in the IV regime.
The second point can be made clear by comparing the MPT
results with those obtained by two other methods, namely the spectral
density approximation (SDA)\cite{MNRR98} and the modified alloy 
analogy (MAA)\cite{RMSRN00pre}. Whereas all three methods do fulfill the
first four moment-sum rules~(\ref{eq:moments}) important for the
strong-coupling limit, the MPT surely
represents a major 
improvement over the other two methods concerning the low-energy
physics. The SDA and the MAA completely neglect the Kondo
physics from the very beginning while the MPT includes, as discussed in
section~\ref{sec:DMFT-modif-pert}, at least qualitatively
the special low-energy properties of the PAM.
Now, if the low-energy
physics have a major influence on ferromagnetism in the IV regime of the
PAM, one would expect a dramatic change of the properties of the
ferromagnetic phase.
However, in all three methods, the ferromagnetic phase turns out to be
very similar, key features as e.\ g.\ the phase diagram, but also the
unusual conduction band polarization are present in all
approximations. This clearly indicates the minor importance of the
special low-energy physics of the PAM when describing ferromagnetism in
the IV regime. The contrary seems to be the case for the Kondo regime:
here the SDA does not give a stable ferromagnetic
phase\cite{MeyerDipl,MNRR98}, whereas within
the MPT, the ferromagnetic solution remains stable in the Kondo regime and
the parameters leading to ferromagnetism
agree very well with the results of reference~\onlinecite{TJF97} if one
ignores the already discussed overestimation of $U_{\rm c}$.
The third point can be clarified by pointing out that for an RKKY
mechanism, the density of the conduction band electrons should have a
major influence on the effective exchange integrals, and therefore also
on phase boundaries. However, as discussed in the context of
figure~\ref{fig:m_n}, the lower phase boundary as function of
$n^{\text{(tot)}}$ is governed by the number of $f$-electrons and is
independent of $n^{(s)}$. This is
the expected behaviour for an intraband mechanism as proposed by us.
It is clear that for the Kondo regime with fixed $n^{(f)}\approx 1$,
the situation is different.
Finally let us discuss the influence of the conduction band
polarization. It is clear that it is decisive for an RKKY mechanism. 
However, our results indicate that the magnitude and the unusual
behaviour of $m^{(s)}$ have no direct effect on the stability of
ferromagnetism in the IV regime (cf.\ figures \ref{fig:m_vef},
\ref{fig:m_n} and \ref{fig:tc_n}). This is a clear indication that an
RKKY mechanism is suppressed here.

To summarize the discussion, we believe that two different mechanism are
to be considered when discussing ferromagnetism in the PAM: In the Kondo
regime, an effective RKKY exchange (interband mechanism) is mainly
responsible for ferromagnetism, in the IV regime, the RKKY exchange is
suppressed and an intraband mechanism similar to the one found for
ferromagnetism in the single-band Hubbard model should be of major
importance. 

From this proposal it follows
that the bandshift (\ref{eq:band-shift}), which is introduced
by the
fourth moment of the spectral density, should also be decisive for
ferromagnetism in the PAM in the IV regime as it was shown to be in the
Hubbard model\cite{PHWN98}. The low-energy physics, however, are vital
in the Kondo regime since the RKKY exchange is believed to origin in the
formation of Kondo screening clouds\cite{TJF97} but less important in
the IV regime. Therefore the shortcomings of the MPT in reproducing the
low-energy scale of the PAM are nonrelevant for the present investigation.

Of course, both the RKKY and the intraband mechanism are always present,
only each respective importance varies. Whereas in the Kondo limit, the
effective $f$-electron itineracy becomes negligible, in the IV regime,
strong-coupling ``Hubbard-like $f$-band'' picture is dominant. This
explains the smooth 
transitions found between the to scenarios (cf.\
figures~\ref{fig:m_vef}, \ref{fig:Uc} and~\ref{fig:tc_ef}).
\begin{figure}[htb]
  \begin{center}
    \epsfig{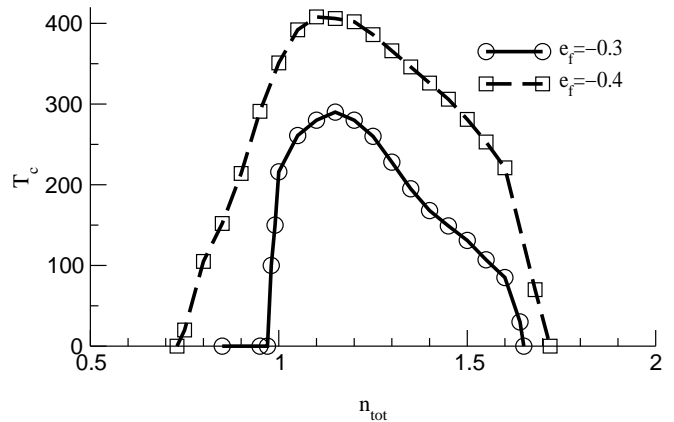}
    \caption{Curie temperatures as function of electron density for
      different $e_f$. The remaining parameters are as in
      figure~\ref{fig:tc_ef}} 
    \label{fig:tc_n}
  \end{center}
\end{figure}

\section{Summary}
\label{sec:sum}
In this paper we have presented a detailed
investigation of ferromagnetism in the intermediate valence regime of
the periodic Anderson model. We applied the dynamical mean-field theory
in combination with the modified perturbation theory. 
We have shortly discussed the quality of the MPT based on a comparison with
exact results for the SIAM and numerically exact results for the
paramagnetic PAM. Although the MPT cannot recover the correct exponential
energy scale for the low-energy physics, the qualitative features of the
low-energy physics emerge properly. And the high-energy features are
reproduced with much higher credibility.

We have established the phase diagram as function of the total electron
density and the position of the $f$-level. Furthermore, an upper bound
of the hybridization strength was found, above which no magnetic ordering
seems to be possible due to local Kondo singlet
formation\cite{MN00b}. 
The critical interaction strength $U_{\rm c}$ was found to vary strongly
when moving from the intermediate valence to the Kondo regime.
The
main contribution to the magnetic moment originates from the localized
$f$-electrons. However, the conduction band is, due to the
hybridization, also polarized.
As
function of electron density, the conduction band polarization changes
sign. For higher electron densities, the conduction-band
polarization is antiparallel to the $f$-level magnetization $m^{(f)}$ as 
one would expect from the Schrieffer-Wolff transformation. For low
densities, this transformation is not applicable. Here,
the conduction band polarization is parallel to $m^{(f)}$. 
The densities of states in the ferromagnetic state show rather
complicated structures. Besides relics of the Kondo resonance and the
dip corresponding to the coherence gap, another gap is present which we
named hybridization gap. This gap is closed when approaching the
paramagnetic state.
Investigating the temperature dependence, we find phase
transitions of second and first order.
We also have presented the Curie temperatures
as function of $f$-level position and total electron density.

Concerning ferromagnetism in the Kondo and Intermediate valence regime,
two different pictures emerge: Whereas in the Kondo regime, an
Kondo-screening induced RKKY exchange leads to the ordering of the
$f$-spins already for relatively low interactions strengths\cite{TJF97},
in the intermediate-valence region, another mechanism becomes more
important: due to the effective
itineracy of the $f$-electrons, the $f$-levels now represent an strongly
correlated narrow band, which by itself can lead to ferromagnetism as
known for the Hubbard model.

\acknowledgements
One of us (D.M.) gratefully acknowledges the support of the
Friedrich-Naumann foundation. This work was further supported by the
Volkswagen foundation.


%
%

%
%


\begin{thebibliography}{10}

\bibitem{hewson}
A.~C. Hewson,
 {\em The Kondo Problem to Heavy Fermions},
 Cambridge University Press 1993.

\bibitem{Czy86}
G.~Czycholl,
 Physics Reports {\bf 143}(5), 277 1986.

\bibitem{GS91}
N.~Grewe and F.~Steglich,
 {\em Heavy Fermions} volume~14 of {\em Handbook on the Physics and Chemistry
  of Rare Earth} page 343,
 Elsevier Science Publishers 1991.

\bibitem{Don77}
S.~Doniach,
 Physica B {\bf 91}, 231 1977.

\bibitem{BFGS87}
R.~Blankenbecler, J.~R. Fulco, W.~Gill, and D.~J. Scalapino,
 Phys. Rev. Lett. {\bf 58}(4), 411 1987.

\bibitem{Jar95}
M.~Jarrell,
 Phys. Rev. B {\bf 51}(12), 7429 1995.

\bibitem{TJF97}
A.N. Tahvildar-Zadeh, M.~Jarrell, and J.K. Freericks,
 Phys. Rev. B {\bf 55}(6), R3332 1997.

\bibitem{Noz98}
P.~Nozi{\`{e}}res,
 Eur. Phys. J. B {\bf 6}, 447 1998.

\bibitem{HMS99}
C.~Huscroft, A.~K. McMahan, and R.~T. Scalettar,
 Phys. Rev. Lett. {\bf 82}(11), 2342 1999.

\bibitem{TJPF99}
A.N. Tahvildar-Zadeh, M.~Jarrell, T.~Pruschke, and J.K. Freericks,
 Phys. Rev. B {\bf 60}(15), 10782 1999.

\bibitem{PBJ00a}
T.~Pruschke, R.~Bulla, and M.~Jarrell,
 Physica B {\bf 281-282}, 47 2000.

\bibitem{VTJK00}
N.~S. Vidhyadhiraja, A.~N. Tahvildar-Zadeh, M.~Jarrell, and H.~R.
  Krishnamurthy,
 Europhys. Lett. {\bf 49}(4), 459 2000.

\bibitem{PBJ00b}
T.~Pruschke, R.~Bulla, and M.~Jarrell,
 Phys. Rev. B {\bf 61}(19), 12799 2000.

\bibitem{MN00b}
D.~Meyer and W.~Nolting,
 Phys. Rev. B {\bf 61}(20), 13465 2000.

\bibitem{TSU97}
H.~Tsunetsugu, M.~Sigrist, and K.~Ueda,
 Rev. Mod. Phys. {\bf 69}(3), 809 1997.

\bibitem{YS93}
T.~Yanagisawa and Y.~Shimoi,
 Phys. Rev. B {\bf 48}(9), 6104 1993.

\bibitem{LM78}
H.J. Leder and B.~M{\"u}hlschlegel,
 Z. Phys. B {\bf 29}, 341 1978.

\bibitem{MNRR98}
D.~Meyer, W.~Nolting, G.G. Reddy, and A.~Ramakanth,
 phys. stat. sol. (b) {\bf 208}, 473 1998.

\bibitem{MN99a}
D.~Meyer and W.~Nolting,
 Physica B {\bf 259}, 918 1999.

\bibitem{MW93}
B.~M{\"o}ller and P.~W{\"o}lfle,
 Phys. Rev. B {\bf 48}(14), 10320 1993.

\bibitem{DS97}
R.~Doradzi\'nski and J.~Spa{\l}ek,
 Phys. Rev. B {\bf 56}(22), R14239 1997.

\bibitem{DS98}
R.~Doradzi\'nski and J.~Spa{\l}ek,
 Phys. Rev. B {\bf 58}(6), 3293 1998.

\bibitem{And99}
F.~B. Anders,
 Phys. Rev. Lett. {\bf 83}(22), 4638 1999.

\bibitem{HC96}
E.~Halvorsen and G.~Czycholl,
 J. Phys.: Condens. Matter {\bf 8}, 1775 1996.

\bibitem{And63}
P.~W. Anderson,
 In F.~Seitz and D.~Turnbull, editors, {\em Solid State Physics: Advances in
  Research and Applications} volume~14 page~99. Academic Press New York 1963.

\bibitem{MV89}
W.~Metzner and D.~Vollhardt,
 Phys. Rev. Lett. {\bf 62}, 324 1989.

\bibitem{PJF95}
T.~Pruschke, M.~Jarrell, and J.~K. Freericks,
 Adv. Phys. {\bf 44}(2), 187 1995.

\bibitem{GKKR96}
A.~Georges, G.~Kotliar, W.~Krauth, and M.~J. Rozenberg,
 Rev. Mod. Phys. {\bf 68}(1), 13 1996.

\bibitem{Geb91}
F.~Gebhard,
 Phys. Rev. B {\bf 44}(3), 992 1991.

\bibitem{Col83}
P.~Coleman,
 Phys. Rev. B {\bf 28}, 5255 1983.

\bibitem{RN83}
N.~Read and D.~M. Newns,
 J. Phys. C {\bf 16}, 3273 1983.

\bibitem{Mue89}
E.~M{\"u}ller-Hartmann,
 Z. Phys. B {\bf 74}, 507 1989.

\bibitem{Ohk91}
F.~J. Ohkawa,
 Phys. Rev. B {\bf 44}(13), 6812 1991.

\bibitem{SC89b}
H.~Schweizer and G.~Czycholl,
 Solid State Commun. {\bf 69}(2), 171 1989.

\bibitem{SC90a}
H.~Schweitzer and G.~Czycholl,
 Solid State Commun. {\bf 74}(8), 735 1990.

\bibitem{KK96}
H.~Kajueter and G.~Kotliar,
 Phys. Rev. Lett. {\bf 77}(1), 131 1996.

\bibitem{PWN97}
M.~Potthoff, T.~Wegner, and W.~Nolting,
 Phys. Rev. B {\bf 55}(24), 16132 1997.

\bibitem{MWPN99}
D.~Meyer, T.~Wegner, M.~Potthoff, and W.~Nolting,
 Physica B {\bf 270}, 225 1999.

\bibitem{MR82}
A~Martin-Rodero, F.~Flores, M.~Baldo, and R.~Pucci,
 Solid State Commun. {\bf 44}, 911 1982.

\bibitem{MR86}
A~Martin-Rodero, E.~Louis, F.~Flores, and C.~Tejedor,
 Phys. Rev. B {\bf 33}, 1814 1986.

\bibitem{Yam75}
K.~Yamada,
 Prog. Theo. Phys. {\bf 53}, 970 1975.

\bibitem{ZH83}
V.~Zlati{\`c} and B.~Horvati{\`c},
 Phys. Rev. B {\bf 28}(12), 6904 1983.

\bibitem{TS99b}
O.~Takagi and T.~Saso,
 J. Phys. Soc. Japan {\bf 68}(9), 2894 1999.

\bibitem{LW60}
J.~M. Luttinger and J.~C. Ward,
 Phys. Rev. {\bf 118}(5), 1417 1960.

\bibitem{Fri56}
J.~Friedel,
 Can. J. Phys. {\bf 34}, 1190 1956.

\bibitem{Lan66}
D.~Langreth,
 Phys. Rev. {\bf 150}(2), 516 1966.

\bibitem{Sas99pre2}
T.~Saso,
 preprint page to appear in Physica B (SCES99) 1999.

\bibitem{And80}
N.~Andrei,
 Phys. Rev. Lett. {\bf 45}(5), 379 1980.

\bibitem{TW83}
A.~M. Tsvelick and P.~B. Wiegmann,
 Adv. Phys. {\bf 32}(4), 453 1983.

\bibitem{HL67}
A.~Harris and R.~Lange,
 Phys. Rev. {\bf 157}(2), 295 1967.

\bibitem{PHWN98}
M.~Potthoff, T.~Herrmann, T.~Wegner, and W.~Nolting,
 phys. stat. sol. (b) {\bf 210}, 199 1998.

\bibitem{Sas97}
T.~Saso,
 J. Phys. Soc. Japan {\bf 66}(4), 1175 1997.

\bibitem{Pruschkeprivate}
T.~Pruschke,
 private communication.

\bibitem{SC89a}
H.~Schweizer and G.~Czycholl,
 Z. Phys. B {\bf 74}, 303 1989.

\bibitem{SC90b}
H.~Schweitzer and G.~Czycholl,
 Z. Phys. B {\bf 79}, 377 1990.

\bibitem{RU85}
T.~M. Rice and K.~Ueda,
 Phys. Rev. Lett. {\bf 55}(9), 995 1985.

\bibitem{NolBd7}
W.~Nolting,
 {\em Viel-Teilchen-Theorie} volume~7 of {\em Grundkurs: Theoretische Physik},
 Friedr. Vieweg \& Sohn Verlagsgesellschft mbH Braunschweig/Wiesbaden 3 edition
  1997.

\bibitem{SW66}
J.~R. Schrieffer and P.~A. Wolff,
 Phys. Rev. {\bf 149}(2), 491 1966.

\bibitem{RMSRN00pre}
G.~G. Reddy, D.~Meyer, S.~Schwieger, A.~Ramakanth, and W.~Nolting,
 submitted.

\bibitem{TJF98}
A.N. Tahvildar-Zadeh, M.~Jarrell, and J.K. Freericks,
 Phys. Rev. Lett. {\bf 80}(23), 5168 1998.

\bibitem{NRM97}
W.~Nolting, S.~Rex, and S.~Mathi Jaya,
 J. Phys.: Condens. Matter {\bf 9}, 1301 1997.

\bibitem{Hub63}
J.~Hubbard,
 Proc. R. Soc. London, Ser. A {\bf 276}, 238 1963.

\bibitem{Gut63}
M.~C. Gutzwiller,
 Phys. Rev. Lett. {\bf 10}(5), 159 1963.

\bibitem{Kan63}
J.~Kanamori,
 Prog. Theor. Phys {\bf 30}, 275 1963.

\bibitem{Ulm98}
M.~Ulmke,
 Eur. Phys. J. B {\bf 1}, 301 1998.

\bibitem{OPK97}
T.~Obermeier, T.~Pruschke, and J.~Keller,
 Phys. Rev. B {\bf 56}(4), R8479 1997.

\bibitem{Vea97}
D.~Vollhardt, N.~Bl{\"u}mer, K.~Held, M.~Kollar, J.~Schlipf, and M.~Ulmke,
 Z. Phys. B {\bf 103}, 283 1997.

\bibitem{MeyerDipl}
D.~Meyer,
 Diploma thesis {H}umboldt-{U}niversit{\"a}t zu {B}erlin May
 1997. (http://tfk.physik.hu-berlin.de/~dmeyer). Although a
 ferromagnetic solution can be found within the SDA for $e_f$ below the
 conduction band, its free energy is then much larger than the one of
 the paramagnetic solution.

\end{thebibliography}
\end{document}